# Electromagnetic and PotentialScattering from a Radially Inhomogeneous Sphere

by John A. Adam and Umaporn Nuntaplook*
Department of Mathematics & Statistics
Old Dominion University, Norfolk, VA 23529

ABSTRACT
Aspects of of plane wave electromagnetic scattering by a radially inhomogeneous sphere is discussed. The vector problem is reduced to two scalar radial 'Schrödinger-like' equations, and a connection with time-independent potential scattering theory is exploited to draw several conclusions about specific refractive index profiles.

1. Introduction

The refractive index $n(r)$ (which may be complex) is a function of the radial coordinate only, and the sphere has radius $a$. For $r > a$, $n(r) \equiv 1$. A time-harmonic dependence of the field quantities, $\exp(-i\omega t)$ is assumed throughout. The governing equation for the electric field $E(r,\theta,\phi)$ is

$$\nabla \times \nabla \times \mathbf{E} - k^2 n^2(r)\mathbf{E} = \mathbf{0}. \tag{1}$$

The wavenumber $k$ is $2\pi/\lambda$, $\lambda$ being the wavelength. As shown in [1], the solution may be found by expanding the electric field in terms of vector spherical harmonics in terms of the so-called transverse electric (TE) and transverse magnetic (TM) modes, respectively:

$$\mathbf{M}_{l,m}(r,\theta,\phi) = \frac{e^{im\phi}}{kr} S_l(r)\mathbf{X}_{l,m}(\theta), \tag{2a}$$

$$\mathbf{N}_{l,m}(r,\theta,\phi) = \frac{e^{im\phi}}{k^2 n^2(r)} \left[ \frac{1}{r}\frac{dT_l(r)}{dr}\mathbf{Y}_{l,m}(\theta) + \frac{T_l(r)}{r^2}\mathbf{Z}_{l,m}(\theta) \right]. \tag{2b}$$

The vector angular functions in equations (2a,b) are defined in a spherical coordinate system as

$$\mathbf{X}_{l,m}(\theta) = \langle 0, i\pi_{l,m}(\theta), -\tau_{l,m}(\theta)\rangle, \tag{3a}$$
$$\mathbf{Y}_{l,m}(\theta) = \langle 0, \tau_{l,m}(\theta), -i\tau_{l,m}(\theta)\rangle, \tag{3b}$$
$$\mathbf{Z}_{l,m}(\theta) = \langle l(l+1)P_l^m(\cos\theta), 0, 0\rangle, \tag{3c}$$

where $P_l^m(\cos\theta)$ is an associated Legendre polynomial of degree $l$ and order $m$. The corresponding scalar angular functions are defined as

$$\pi_{l,m}(\theta) = \frac{m}{\sin\theta}P_l^m(\cos\theta), \tag{4a}$$
$$\tau_{l,m}(\theta) = \frac{dP_l^m(\cos\theta)}{d\theta}. \tag{4b}$$

The functions $S_l(r)$ and $T_l(r)$ are called the radial Debye potentials, and they respectively satisfy the equations

$$\frac{d^2 S_l(r)}{dr^2} + \left[k^2 n^2(r) - \frac{l(l+1)}{r^2}\right] S_l(r) = 0, \tag{5a}$$

$$\frac{d^2 T_l(r)}{dr^2} - \left(\frac{2}{n(r)}\frac{dn(r)}{dr}\right)\frac{dT_l(r)}{dr} + \left[k^2 n^2(r) - \frac{l(l+1)}{r^2}\right] T_l(r) = 0. \tag{5b}$$

In addition to the appropriate matching conditions at $r = a$ these potentials must also satisfy the boundary conditions

$S_l(0) = 0$ and $T_l(0) = 0$. Equation (5b) may be rewritten in terms of the dependent variable $U_l(r)$, where $T_l(r) = n(r)U_l(r)$ to become

$$\frac{d^2 U_l(r)}{dr^2} + \left[ k^2 n^2(r) - n(r) \frac{d^2}{dr^2}\left(\frac{1}{n(r)}\right) - \frac{l(l+1)}{r^2} \right] U_l(r) = 0. \tag{6}$$

Provided that $n(0) \neq 0$, $U_l(0) = 0$. Both equations (5a) and (6) may be placed in the form of the canonical time-independent Schrödinger equation, namely

$$\frac{d^2 S_l(r)}{dr^2} + \left[ k^2 - V_S(r) - \frac{l(l+1)}{r^2} \right] S_l(r) = 0, \tag{7a}$$

$$\frac{d^2 U_l(r)}{dr^2} + \left[ k^2 - V_U(r) - \frac{l(l+1)}{r^2} \right] U_l(r) = 0, \tag{7b}$$

where the $k$-dependent 'scattering potentials' $V_S(r)$ and $V_U(r)$ are defined in $[0, a]$ as

$$V_S(r) = k^2[1 - n^2(r)], \tag{8a}$$

$$V_U(r) = k^2\left[ 1 - n^2(r) + \frac{n(r)}{k^2} \frac{d^2}{dr^2}\left(\frac{1}{n(r)}\right) \right]. \tag{8b}$$

for the TE and TM modes respectively (the potentials are both identically zero for $r > a$). These potentials are identical for the case of a uniform refractive index. $V_U(r)$ will be regarded as a small perturbation of the potential $V_S(r)$, so we also define

$$\varepsilon(r) \equiv V_U(r) - V_S(r) = n(r) \frac{d^2}{dr^2}\left(\frac{1}{n(r)}\right). \tag{9}$$

It is a standard result for potentials vanishing sufficiently fast at infinity [2–4] that as $r \to \infty$

$$S_l(r) \sim \sin\left(r - \frac{\pi l}{2} + \delta_l^S(k)\right), \tag{10a}$$

$$U_l(r) \sim \sin\left(r - \frac{\pi l}{2} + \delta_l^U(k)\right). \tag{10b}$$

Here $\delta_l^S(k)$ and $\delta_l^U(k)$ are the phase shifts induced by each potential respectively. Multiplying equations (7a) and (7b) by $U_l(r)$ and $S_l(r)$ respectively, subtracting and integrating we obtain

$$U_l(r)\frac{dS_l(r)}{dr} - S_l(r)\frac{dU_l(r)}{dr} = -\int_0^r \varepsilon(\eta) S_l(\eta) U_l(\eta) d\eta. \tag{11}$$

Utilizing the asymptotic expressions in (10), we have, in the limit as $r \to \infty$,

$$k \sin[\delta_l^U(k) - \delta_l^S(k)] = -\int_0^\infty \varepsilon(r) S_l(r) U_l(r) dr = -\int_0^{ka} \varepsilon(r) S_l(r) U_l(r) dr, \tag{12}$$

since $n(r)$ is constant for $r > ka$ (or $r > a$). Thus far this equation is exact. If we now consider $\varepsilon(r)$ to be sufficiently small that $U_l(r) \approx S_l(r)$, then $|\delta_l^U(k) - \delta_l^S(k)| \ll 1$ and we have the relation

$$\delta_l^U(k) \approx \delta_l^S(k) \pm \frac{1}{k} \int_0^{ka} \varepsilon(r)[S_l(r)]^2 dr. \tag{13}$$

Whether $\delta_l^U(k) > \delta_l^S(k)$ or not clearly depends on the concavity of $n(r)$. A further approximation can be made if the scattering potential $V_S(r)$ is constant (specifically, $V_S = k^2(1 - N^2)$ for $n = N$, $r \leq a$), for then the solution for equation (7a) can be expressed in terms of a Riccati-Bessel function of the first kind, i.e.

$$S_l(r) = \left(\frac{\pi N k r}{2}\right)^{1/2} J_{l+1/2}(Nkr). \tag{14}$$

Then we have that (check!)

$$\delta_l^U(k) \approx \delta_l^S(k) \pm \frac{\pi N}{2} \int_0^a \left\{ n(r) \frac{d^2}{dr^2}\left(\frac{1}{n(r)}\right) \right\} [J_{l+1/2}(Nkr)]^2 r\, dr \equiv \delta_l^S(k) \pm \frac{\pi N}{2} \mathcal{I}(a). \tag{15}$$

In the case of a small perturbation about $V_S = 0$, i.e. for which $n = N = 1$, the term $\delta_l^S(k)$ in equation (15) is zero, and the resulting approximation for $\delta_l^U(k)$ is related to the first Born approximation in quantum scattering theory [5]. In particular, if $\varepsilon(r) = Dr^{-s}$, $D$ being some constant, a closed form solution for $\mathcal{I}$ can be found as $a \to \infty$ [4], namely

$$\mathcal{I}(\infty) = \int_0^\infty [J_{l+1/2}(Nkr)]^2 r^{1-s} dr = \frac{1}{2}\left(\frac{Nk}{2}\right)^{s-2} \frac{\Gamma(s-1)\Gamma\left(l - \frac{1}{2}s + \frac{3}{2}\right)}{\left[\Gamma\left(\frac{1}{2}s\right)\right]^2 \Gamma\left(l + \frac{1}{2}s + \frac{1}{2}\right)}, \tag{16}$$

provided $s > 1$ and $2l > s - 3$. The question may be asked: what $n(r)$ profiles give rise to $\varepsilon(r) = Dr^{-s}$ (where $D > 0$)? Writing $p(r) = [n(r)]^{-1}$ we are led to consider solutions of the equation

$$r^s \frac{d^2 p(r)}{dr^2} - Dp(r) = 0. \tag{17}$$

The general solution to this equation may be expressed in terms of modified Bessel functions, but we do not pursue this direction here.

A Liouville transformation

As defined in equations (8a) and (8b), the 'potentials' $V_S(r)$ and $V_U(r)$ are also $k$-dependent, which is not the case in potential scattering theory [3]. This has an important consequence: unlike the quantum mechanical case, here pure 'bound state' solutions, that is, real square-integrable solutions corresponding to $k^2 < 0$ ($\operatorname{Im} k > 0$) do not exist. This can readily be proven [5,6] for the TE mode (equation (7a)) that

$$\int_0^\infty \left[ \left|\frac{dS_l(r)}{dr}\right|^2 + \frac{l(l+1)}{r^2}|S_l(r)|^2 \right] dr = k^2 \int_0^\infty n^2(r)|S_l(r)|^2 dr. \tag{18}$$

This cannot be satisfied for $k^2 < 0$ for a real and positive refractive index $n(r)$. In [7] the corresponding result is established from equation (7b) for $U_l(r)$. Furthermore, a Liouville transformation may be used to define a new $k$-independent potential [5]. Using the following simultaneous changes of independent and dependent variables in equation (5a)

$$r \to \rho : \rho(r) = \int_0^r n(s) ds, \tag{19a}$$

$$u_l \to \psi_l : \psi_l(\rho) = (n(r))^{1/2} u_l(r). \tag{19b}$$

Clearly $n(r)$ must be integrable and non-negative (in naturally-occurring circumstances $n \geq 1$ and $n(r) = 1$ for $r > a$); also $\rho(0) = 0$. It is easy to establish the following results:

(i) $\rho(r) = \rho_0 + r - a$, $r \geq a$, where $\rho_a = \int_0^a n(s) ds$;

(ii) $\rho(r) \sim r$, $r \to \infty$;

(iii) $r(\rho) = \int_0^\rho \frac{ds}{v(s)}$, where $v(\rho) = n(r(\rho))$.

Furthermore, by applying (19a) and (19b) to equation (7a) we find that

$$\left[\frac{d^2}{d\rho^2} - \frac{l(l+1)}{R^2(\rho)} + k^2\right]\psi_l(r) = V(\rho)\psi_l(\rho), \qquad (20)$$

where

$$R(\rho) = v(\rho)r(\rho) \sim n(0)\rho, \ \rho \to 0, \text{ and } V(\rho) = [v(\rho)]^{-1/2}\frac{d^2}{d\rho^2}[v(\rho)]^{1/2}. \qquad (21)$$

Clearly $v(\rho)$ should be at least twice differentiable. Now the new 'potential' $V(\rho)$ is independent of the wavenumber $k$. Note also that $V(\rho) = 0$ for $\rho > \rho_a$. It is of interest to determine the 'shape' of the potential $V(\rho)$ by inverting $\rho(r)$ for various choices of physical $n(r)$ profiles for $r \in [0,a]$ (with $n(0) = n_0$, $n(a) = n_a$ and $n(r) = 1$ for $r > a$). In what follows only the non-zero potential shapes with be stated (corresponding to $\rho \in [0,\rho_a]$. Thus [5] for

$$n(r) = n_a\left[1 - c^2\left(\frac{r-a}{a}\right)^2\right]^{-1}; \ V(\rho) = \frac{c^2}{n_a^2} > 0, \qquad (22a)$$

where $c$ is a real constant, i.e. the potential is a spherical barrier. For the profile [8]

$$n(r) = (A + Br)^{-1}, \ A = n_0^{-1}, \ B = \frac{n_0 - n_a}{an_0n_a}; \ V(\rho) = \frac{B^2}{4} > 0, \qquad (22b)$$

also a barrier. For the important Maxwell Fish-Eye profile [9],

$$n(r) = n_0(1 + Br^2)^{-1}, \ B = \frac{n_0 - n_a}{a^2n_a}; \ V(\rho) = -\frac{B}{n_0^2}. \qquad (22c)$$

In this case, the new potential is a spherical well or barrier as $n_0 > n_a$ or $n_0 < n_a$ respectively. In the latter case the singularity occurring in $n(r)$ is moot since it arises for $r > a$. In all the other cases investigated thus far [10], including $n(r) = n_0 \exp(-\alpha r); \ n_0 \cos \alpha r$ and $n_0 \cosh \alpha r$, the potentials $V(\rho)$ are rather complicated functions, and there are no significant advantages to using the Liouville transformation in these cases. It is therefore of interest to examine what profiles $n(r)$ give rise to constant potentials $V(\rho)$. In equation (21) let $y(\rho) = [v(\rho)]^{1/2}$ and $V(\rho) = V_0$, where $V_0$ is a constant of either sign. Then it follows that

$$\frac{d^2y}{d\rho^2} - V_0y = 0, \qquad (23)$$

the general solution being expressible in terms of real or complex exponential functions as $V_0 > 0$ (potential barrier) or $V_0 < 0$ (potential well) respectively. In $r$-space, $V_0 < 0$ corresponds to a constant refractive index $n = N = (1 + |V_0|k^{-2})^{1/2} > 1$, so we proceed with this physically realistic case. Writing the general solution of (23) as

$$y(\rho) = C\cos\left(|V_0|^{1/2}\rho + \eta\right), \qquad (24)$$

where $C$ and $\eta$ are constants, it follows that

$$r(\rho) = \int_0^\rho \frac{ds}{v(s)} = \left(C^2|V_0|^{1/2}\right)^{-1}\left[\tan\left(|V_0|^{1/2}\rho + \eta\right) - \tan\eta\right]. \qquad (25)$$

This can be inverted to yield

$$\rho(r) = \int_0^r n(s)ds = |V_0|^{-1/2}\left\{\arctan\left[C^2|V_0|^{1/2}r + \tan\eta\right] - \eta\right\}. \qquad (26)$$

Therefore

$$n(r) = \rho'(r) = \frac{C}{1 + [Br + \tan \eta]^2}, \tag{27a}$$

where $C = n_0 \sec^2 \eta$ and $\eta$ can be determined from the requirement that $n(a) = n_a$. This is a generalization of the Maxwell Fish-Eye profile in equation (22c). The corresponding result for $V_0 > 0$ is

$$n(r) = \frac{C}{1 - [Br + \tanh \eta]^2}. \tag{27b}$$

Note that in this case a singularity exists for $r > 0$ at $r = B^{-1}(1 - \tanh \eta)$.

References

[1] B.R. Johnson, Theory of morphology-dependent resonances: shape resonances and width formulas, J. Opt. Soc. Am. A10 (1993) 343-352.
[2] L.I. Schiff, Quantum Mechanics, 3rd edition (1968), New York, McGraw-Hill.
[3] V. de Alfaro and T. Regge, Potential Scattering (1965), North-Holland Publishing Company, Amsterdam.
[4] N.F. Mott and H.S. W. Massey, The Theory of Atomic Collisons, 3rd edition (1965), Clarendon Press, Oxford.
[5] C. Eftimiu, Direct and inverse scattering by a sphere of variable index of refraction, J. Math. Phys. 23 (1982) 2140-2146.
[6] J. A. Adam, (to appear): 'Rainbows' in homogeneous and radially inhomogeneous spheres: connections with ray, wave and potential scattering theory,
   Advances in Interdisciplinary Mathematical Research: Applications to Engineering, Physical and Life Sciences,
   Springer Proceedings in Mathematics & Statistics, Vol. 37. Ed. Bourama Toni, Springer, 2013.
[7] Eftimiu, C., 1985: Inverse electromagnetic scattering for radially inhomogeneous dielectric spheres, in Inverse methods in electromagnetic imaging;
   Proceedings of the NATO Advanced Research Workshop, Bad Windsheim, West Germany, September 18-24 (1983), Part 1 (A85-48926 24-70).
   Dordrecht, D. Reidel Publishing Company.
[8] J. A. Adam and P. Laven, P.,On rainbows from inhomogeneous transparent spheres: a ray-theoretic approach, Appl. Opt. 46 (2007) 922–929.
[9] U. Leonhardt and T. Philbin, Geometry and Light: The Science of Invisibility (2010), New York: Dover Publications.
[10] J. A. Adam and U. Nuntaplook, in preparation.